\documentclass[twocolumn,showpacs,preprintnumbers,amsmath,amssymb]{revtex4}

\usepackage{graphicx}
\usepackage{amssymb}
\usepackage{dcolumn}% Align table columns on decimal point

\newcommand {\be}{\begin{equation}}
\newcommand {\ee}{\end{equation}}
\newcommand {\bea}{\begin{eqnarray}}
\newcommand {\eea}{\end{eqnarray}}
\newcommand {\FIG}[1]{Fig. \ref{#1}}
\newcommand {\EQ}[1]{Eq. (\ref{#1})}
\newcommand {\REF}[1]{Ref. \cite{#1}}

\newcommand {\PRB}[1]{Phys. Rev. B {\bf {#1}}}
\newcommand {\PRL}[1]{Phys. Rev. Lett. {\bf {#1}}}

\newcommand {\NAT}[1]{Nature {\bf {#1}}}
\newcommand {\JSP}[1]{J. Stat. Phys. {\bf {#1}}}
\newcommand {\JPA}[1]{J. Phys. A: math. gen. {\bf {#1}}}

\begin{document}
\title{Diffusive Capture Process on Complex Networks}

\author{Sungmin Lee}
\author{Soon-Hyung Yook}
\email{syook@khu.ac.kr}
\author{Yup Kim}
\email{ykim@khu.ac.kr}
\affiliation{Department of Physics and
Research Institute for Basic Sciences, Kyung Hee University, Seoul
130-701, Korea}
\date{\today}

\begin{abstract}
We study the dynamical properties of a diffusing lamb
captured by a diffusing lion on the complex networks with various
sizes of $N$.
We find that the life time $\left<T\right>$ of a lamb scales
as $\left<T\right>\sim N$ and the survival probability
$S(N\rightarrow \infty,t)$
becomes finite on scale-free networks with degree exponent $\gamma>3$.
However, $S(N,t)$ for $\gamma<3$ has a
long-living tail on tree-structured scale-free networks
and decays exponentially on looped scale-free networks.
It suggests that the second moment of degree distribution
$\left<k^2\right>$ is the relevant factor for the dynamical properties
in diffusive capture process.
We numerically find that the normalized number of capture events at
a node with degree $k$,
$n(k)$, decreases as $n(k)\sim k^{-\sigma}$.
When $\gamma<3$, $n(k)$ still increases anomalously for $k\approx k_{max}$.
We analytically show that $n(k)$ satisfies the relation
$n(k)\sim k^2P(k)$ and the total number of capture events
$N_{tot}$ is proportional to $\left<k^2\right>$, which causes the
$\gamma$ dependent behavior of $S(N,t)$ and $\left<T\right>$.
\end{abstract}
\maketitle

\section{INTRODUCTION}

The diffusion-controlled reactions, in which diffusing particles
are immediately converted to a product if a pair of them meets
together, have many physical applications. Examples include
electron trapping and recombination, exciton fusion, wetting,
melting, and commensurate-incommensurate transitions
\cite{agranovich,parus89,ben-avraham88,fisher84,huse84,
lipowsky88,redner-book}. Among these examples, dynamical properties
of wetting, melting, and commensurate-incommensurate transition
are known to be related to the diffusive capture process
\cite{ben-avraham88,fisher84,huse84,lipowsky88,redner-book}, whose
kinetics can be simplified by diffusing preys-predators model
\cite{krapivsky,redner}.

In general, the dynamical properties of diffusion-controlled
reactions are closely related to the first-passage phenomena of
random walks (RWs), because the reaction occurs when diffusing
particles first meet a trap or an absorbing boundary
\cite{redner-book,krapivsky,redner}. Especially, the diffusive
capture process can be mapped into coupled RWs \cite{bramson}, in
which $n$ predator random walkers (lions or moving trap) stalk a
prey random walkers (a lamb or a moving particle).
Since the first-passage probability can
be easily obtained from the survival probability
\cite{redner-book}, the survival probability of a diffusing
particle or a lamb is important to understand long time
characteristics of the diffusive capture process
\cite{krapivsky,redner}. The survival probability of a lamb has
been well investigated in $d$-dimensional hypercubic
lattice. The diffusive capture process in the spatial dimension $d
> 2$ is unsuccessful and the survival probability is finite for
any $n$. In $d=2$, the capture is successful but not efficient,
i.e., the life time of a prey random walker is infinite and the
survival probability shows a logarithmic decay \cite{aspects-rw}.
The predators efficiently capture a prey for spatial dimension
$d<2$. On a $d$-dimensional regular lattice, the survival
probability $S(t)$ depends on $d$ as follows
\cite{krapivsky,redner,bramson,Feller,Weiss}
\begin{equation}
S(t) \sim \left\{
\begin{array}{lll}
t^{-\beta} & for & d < 2\\
(\ln{t})^{-n} & for & d = 2\\
\mbox{finite} & for & d > 2,
\end{array}\right.
\end{equation}
where the exponent $\beta$ varies as $n$ and the ratio of
diffusion constant of walks.

There have been increasing interests on complex networks, since
many structures of physically interacting systems are shown to
form nontrivial complex networks \cite{Jeong01, DM, www_ajb,
www_broder, internet_fal, internet_yook}. The early studies on
complex networks mainly focused on their topological properties.
Recently, the physical systems whose elements interact along the
links in complex networks and the effects of the underlying
topologies on the dynamical properties of the system have drawn
much attention. Many dynamical systems on complex networks show
rich behaviors which are far from the mean-field expectations
\cite{Critical, Ferromagnetic, random_break, Percolation}.
Recently, RWs on complex networks have been studied to show that
the dynamical properties of RWs are closely related
to the topology of underlying networks \cite{Noh92,
noh_tree_loop}. This implies that the first-passage property of
the diffusive capture process is also affected by the topology of
underlying network. Moreover, the several studies on RWs have
provided very interesting application to searching information
over the Internet \cite{Huberman, our_llonnet}. For example, the
major searching engines such as Google and YAHOO use general
random walking robots along the links between hyper-texts to
collect information of each web page \cite{ves_book}. Some
peer-to-peer (P2P) network protocols also exploit the dynamical
properties of RWs on complex networks \cite{Huberman}. Especially,
the searching algorithm for P2P network can be mapped to the
system of a diffusing particle to find an immobile absorbing
particle (static trap). Our recent study \cite{our_llonnet} has
also found that the searching algorithm inspired by the diffusive
capture process can dramatically reduce the traffic over the
Internet with scalable searching time. The diffusive capture
process can be also applied to investigating condensed phase of
conserved-mass aggregation model on scale-free networks (SFNs)
with $2 < \gamma \leq3$ \cite{cmaonsfn}. Therefore, it is
theoretically and technologically very interesting to study
the effect of nontrivial complex networks on the diffusive
capture process.

In order to understand the relationship between the non-trivial
topology of underlying network and the diffusive capture process on
it, we study the random walking lamb and lions on SFNs.
The SFNs are the networks whose degree distribution follows
$P(k)\sim k^{-\gamma}$.
The degree exponent $\gamma$ of the most natural and technological
networks is in the range $2 < \gamma\le 3$ \cite{DM,huberman_pk}.
Moreover, the probability that a random walker visits a node
has been shown to be proportional to the degree $k$ of the node
\cite{Noh92}. So, the probability that two random walking
particles meet at a node is proportional to $k^2$. Therefore, we
expect that the dynamical properties of the diffusive capture process on
the SFNs with the degree exponent $2<\gamma \leq 3$ are different
from those on the SFNs with $\gamma > 3$, because $\left<k^2\right>$
diverges for $\gamma<3$. To verify our expectations, we study the
diffusive capture process on the SFNs. More recently, Noh et al.
\cite{noh_tree_loop} also showed that the return probability of a
random walker on the looped SFNs (LSFNs) is completely different
from that on the tree-structured scale-free networks
(TSFNs). Therefore, the diffusive capture process on TSFNs
is also investigated.

This paper is organized as follows: In Sec. II, we present a
method of generating TSFNs and dynamical properties of
the diffusive capture process on TSFNs. The results on LSFNs are shown
in Sec. III. In Sec. IV, we discuss the relation between degrees
of SFNs and capture events. Finally we summarize our results in Sec. V.

\section{DIFFUSIVE CAPTURE PROCESS ON TSFNs}

All the models are defined on a graph $Gr=\{N,K\}$, where $N$ is
the number of nodes in the networks and $K$ is the number of
degrees with the average degree $\left<k\right>=2K/N$. Networks
with the sizes $N=10^3 \sim 10^6$ are generated for each $\gamma$.

To make tree structure, the total number of links in networks with
the size $N$ is constrained to be $K_T=2(N-1)$. So we use the following
method to make TSFNs. Select $N$ integer
random numbers $\{k_1, k_2,\cdots,k_N\}$ whose probability distribution
satisfies $f(k) \sim k^{-\gamma}$. Let $K_0 = \sum^{N}_{i=1} k_i$.
If $K_T>K_0$ ($K_T<K_0$), then choose $k_j$ randomly from the set
$\{k_i\}$ and decrease (increase) by 1 as $k_j \rightarrow k_j +1$
($k_j \rightarrow k_j -1$).
If the resulting $k_j$ becomes $0$, the above process is rejected and
take the same process again. This process is repeated until $K_T=K_0$ is
fulfilled. Sort the final set $\{k_1, k_2, \cdots, k_N\}$ with $K_T
=\sum_{i=1}^N k_i$ to satisfy $k_1 > k_2 > \cdots > k_N$ and set 
$k_i$ to be the degree of the node
$i$. Now link the nodes $2$, $3$, $\cdots$, $k_1 +1$ to the node
$1$ to make the degree of the node $1$ be $k_1$. Next, link the nodes
$k_1 +2$, $k_1 +3$, $\cdots$, $k_1 +k_2$ to the node $2$ to make
the degree of the node $2$ be $k_2$. Repeat the same
processes until no isolated node remains.

\begin{figure}
\includegraphics[width=8.5cm]{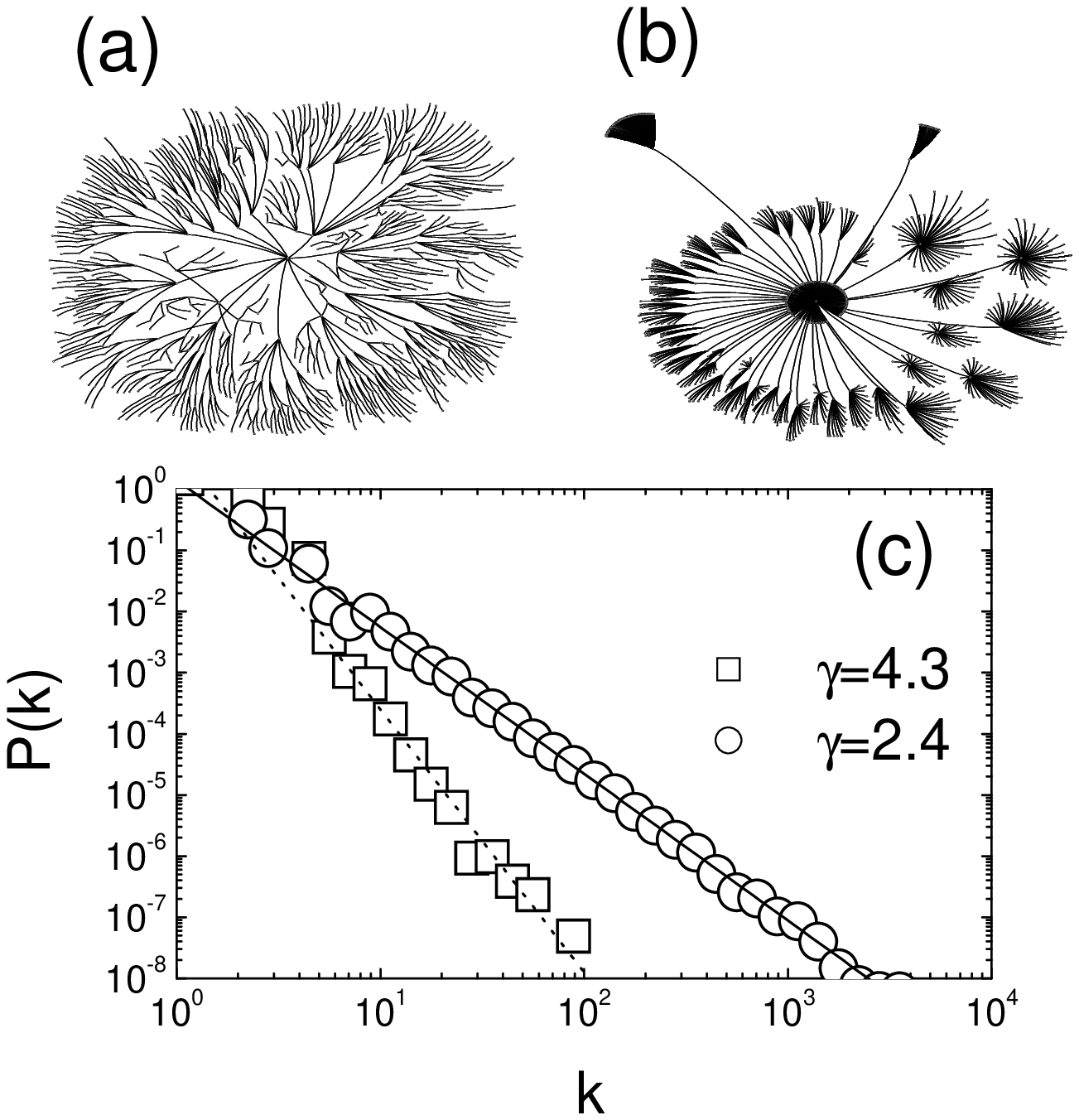}
\caption{Schematic structures of TSFNs with $\gamma=4.3$ (a), and $2.4$ (b)
constructed by the method introduced in Sec. II.
The plot of the degree distribution of the TSFNs with $\gamma=2.4$, and $3.0$ (c).
Network size is $N=10^5$. The degree distribution $P(k) \sim k^{-\gamma}$
of the constructed networks follows the original
distribution $f(k) \sim k^{-\gamma}$ well.} \label{PK}
\end{figure}

Figs. \ref{PK}(a) and \ref{PK}(b) show snapshots of typical networks
generated by the above procedure with 1000 nodes.  \FIG{PK}(c)
shows the degree distribution of TSFNs constructed with $\gamma=2.4$
and $4.3$. To generate TSFNs we use $N=10^5$. The data show a good
agreement between the exponents $\gamma$ obtained from the degree distribution,
$P(k) \sim k^{-\gamma}$, of the constructed network (symbols) and $\gamma$
of the original distribution $f(k) \sim k^{-\gamma}$ (solid and dashed lines).
As shown in Figs. \ref{PK}(a) and \ref{PK}(b), the topological structure of
TSFNs with $\gamma>3$ is completely different from that with $\gamma<3$.
We find relatively small degree fluctuation in the
networks with $\gamma>3$ (see \FIG{PK} (a)).
The small fluctuation causes almost homogeneous
topological properties. Thus, we expect that the dynamical properties on the
networks with $\gamma>3$ follow the mean-field expectations, i.e.,
the non-zero survival probabilities of a lamb in $N\rightarrow \infty$
limit. However, for network with $\gamma<3$, there exist many short
branches and a big hub having most links. The dynamical
properties of a given system on this kind of network can be
dominated by this hub. Therefore, as we shall show later,
the lamb can easily hide from the lions by moving to a different branch
through the hub.

Now, let's discuss the dynamical properties of the diffusive capture
process on the constructed TSFNs. Initially the lamb and $n$ lions
are randomly distributed to the nodes on the networks. In the
simulations, we consider two updating rules to determine the
position of the random walking lamb and lions: 1) parallel update
and 2) random update. In the parallel update rule, we move all
lions and a lamb at the same time, then check if the lamb meets
the lion(s). In this algorithm, the lamb and the lions should be
initially separated by $2m$ $(m=1,2,\cdots )$ steps. Otherwise,
the lamb never meets the lion moving on TSFNs. In the random update
rule, we choose one among the lamb or lions at random. The
randomly chosen lamb or lion takes a random walk and then check if
the lamb meets the lion. The unit time is defined by $(n+1)$ such
random movements. We find that there is no significant difference
between dynamical behaviors for each update rule. We also find that
the dynamical behaviors are not varied critically if $n$ is
varied. Therefore, we will mainly show the simulation results for
1 lamb and 10 lions in this article. All the numerical results are
obtained by averaging over $100$ network realizations and $1000$
histories for each network realization.

\begin{figure}
\includegraphics[width=8.5cm]{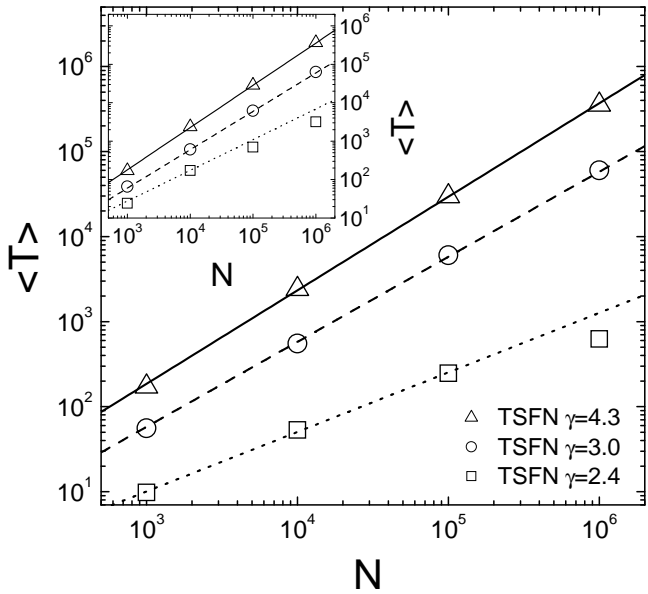}
\caption{ Average life time $\left< T \right>$ on TSFNs
with $\gamma=4.3$, $3.0$ and $2.4$. The lines represent the best
fits of $\left<T\right> \sim N^{\alpha}$ to data.
The obtained exponents are $\alpha\simeq 1.1$ (solid line) for $\gamma=4.3$,
$\alpha=1.0$ (dashed line) for $\gamma=3.0$ and $\alpha=0.7$ (dotted line)
for $\gamma=2.4$. For $\gamma=2.4$ $\left< T \right>$ substantially deviates
from $\left<T\right> \sim N^{0.7}$ line when $N=10^6$.}
\label{TSFN_T}
\end{figure}

\FIG{TSFN_T} shows the plot of the average life time
$\left<T\right>$ on TSFNs with various $\gamma$ against
$N$ using the random update rule. $\left<T\right>$ increases with
$N$ for any $\gamma$. The lines represent the best fits of the relation
$\left<T\right> \sim N^{\alpha}$ to the data.
From the fits, we find that $\left<T\right>$ follows
the power-law $\left<T\right>\sim N^\alpha$ with $\alpha=1$
for $\gamma\ge 3$, and $\alpha=0.7$ for $\gamma=2.4$ (dotted line).
But, as shown in the \FIG{TSFN_T}, $\left<T\right>$ for $\gamma=2.4$
deviates substantially from the power-law relation $\left<T\right>\sim
N^{0.7}$ as increasing $N$ (square).
The inset shows the results obtained by using the parallel update
rule. The data obtained from the parallel update rule show nearly the
same dynamical behaviors as those obtained from the random
update rule.

\begin{figure}
\includegraphics[width=8.5cm]{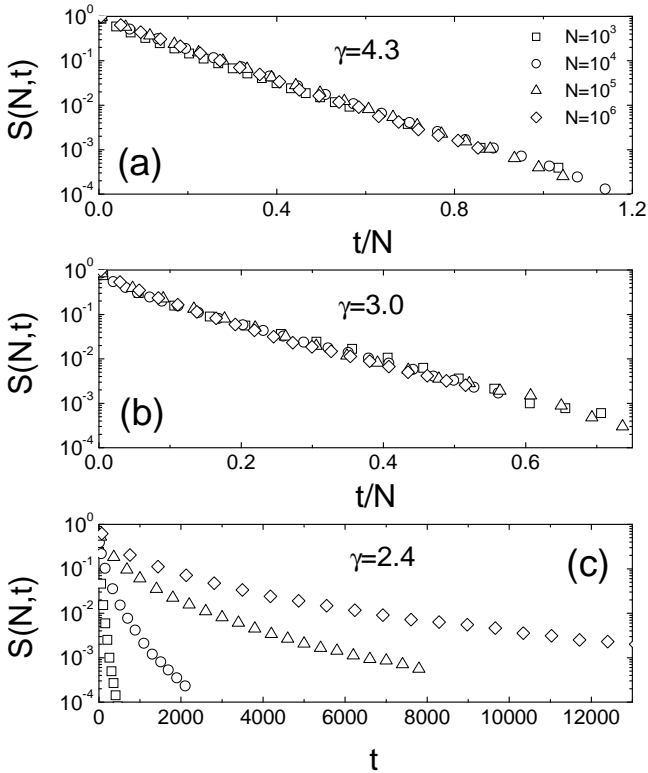}
\caption{The scaled plots of the survival probability $S(N,t)$ on
TSFNs with $\gamma=4.3$ (a), $3.0$ (b) and the plot of $S(N,t)$ for
$\gamma=2.4$ (c). For the scaling plots, $\alpha=1$ which is obtained
from the fits for $\gamma>3$ in \FIG{TSFN_T}
is used.}
\label{TSFN_ST}
\end{figure}

We also measure the survival probability on TSFNs with various $\gamma$
and $N$. \FIG{TSFN_ST} shows $S(N,t)$ for $\gamma=4.3$ (a), $3.0$
(b) and $2.4$ (c) from the random update rule.
As is shown in \FIG{TSFN_ST}, the survival probability for
$\gamma \ge 3$ exponentially decays as $S(N,t) \sim \exp(-t/\tau(N))$,
however, $S(N,t)$ for $\gamma < 3$ manifests a substantial
deviation from an exponential decay.
If we assume that $S(N,t)= S_o e^{-t/\tau}$, then we obtain
$\left<T\right> \sim \tau(N)$ from the relation
$\left<T\right> = \int^{\infty}_0 t (- dS(N,t)/dt) dt$.
Therefore, the characteristic time $\tau$ of $S(N,t)$ scales in
the same manner as $\left<T\right>$, and the scaling $t/N$
collapses $S(N,t)$'s for various $N$ into one curve when 
$\gamma \ge 3$ [see \FIG{TSFN_ST}(a),
(b)]. The relation $\tau(N)\sim N$ implies that $S(N,t)$ becomes
a nonzero constant in the limit $N\rightarrow \infty$.
Therefore, the exponentially decaying $S(N,t)$ for $\gamma > 3$
can be regarded as the finite size effect. However, $S(N,t)$ for
TSFN with $\gamma=2.4$ shows the long-living tail [\FIG{TSFN_ST}(c)]
which decreases more slowly than exponential decay.
Hence, we expect
\begin{equation}
\label{ST_TSFN}
\lim_{t\rightarrow \infty}\lim_{N \rightarrow \infty} S(N,t)= \begin{cases}
\mbox{long-living tail} & (\gamma < 3)\\
\mbox{finite}  & (\gamma \ge 3) \;\;.
 \end{cases}
\end{equation}
The origin of $\gamma$ dependent behaviors of $S(N,t)$ will be
discussed in Sec. IV.

\begin{figure}
\includegraphics[width=8.5cm]{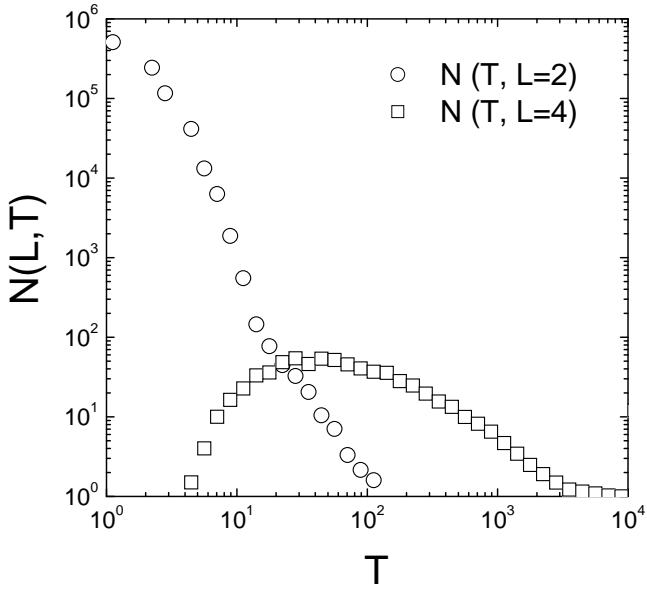}
\caption{ Plot of $N(L,T)$ on TSFN with
$\gamma=2.4$ and $N=10^4$. The initial separation $l$ of the lamb
and the lion is $2$. $L$ is the maximum separation
between the lamb and the lion during random walks (symbols). }
\label{ft}
\end{figure}

We now discuss the origin of the long-living
tail of $S(N,t)$ for $\gamma=2.4$. One possible explanation of the
slowly decaying tail of $S(N,t)$ can be understood from the
topological properties of TSFNs. As shown in \FIG{PK} (b), the
biggest hub in the TSFN with $\gamma<3$ has many short branches.
Hence, if the lamb hides from the lions by moving into a different
branch, then the lion should visit almost all branches to search
for a hidden lamb. Thus, the hidden lamb can live longer than the
one in the same branch with the lion. Therefore, $S(N,t)$ on
TSFN with $\gamma<3$ can have the long-living tail. In order to
verify this argument, we count the
number of capture events $N(L,T)$ with the life time $T$ among $10^6$
different histories on TSFN with $\gamma=2.4$ (\FIG{ft})
by using the random update rule.
At the same time we also measure the maximum separation $L$ between the
lamb and the lion during random walks.
Initially we set the shortest path length $l$ between the
lamb and lion to be 2.
We use the TSFN whose average diameter ${\bar d}$
is ${\bar d}\simeq 4$. If $L\simeq {\bar d}$, then
the lamb and the lion have once been in different branches
during random walks. 
The data in \FIG{ft} explicitly shows that
$N(L,t)$ for $L=2$ decays fast (the maximum life time of a lamb is
$T_{max}\simeq 10^2$). In contrast, $N(L,t)$ for $L=4$ shows the
long-living tail up to $T_{max}\simeq 10^4$.
From the relation $S(t)\propto -\int_t \sum_L N(L,t)dt$,
one can easily see that the long-living tail of $S(t)$ is
caused by the diffusive capture process with $L=4$.

\section{Diffusive capture process on LSFNs}

\begin{figure}
\includegraphics[width=8.5cm]{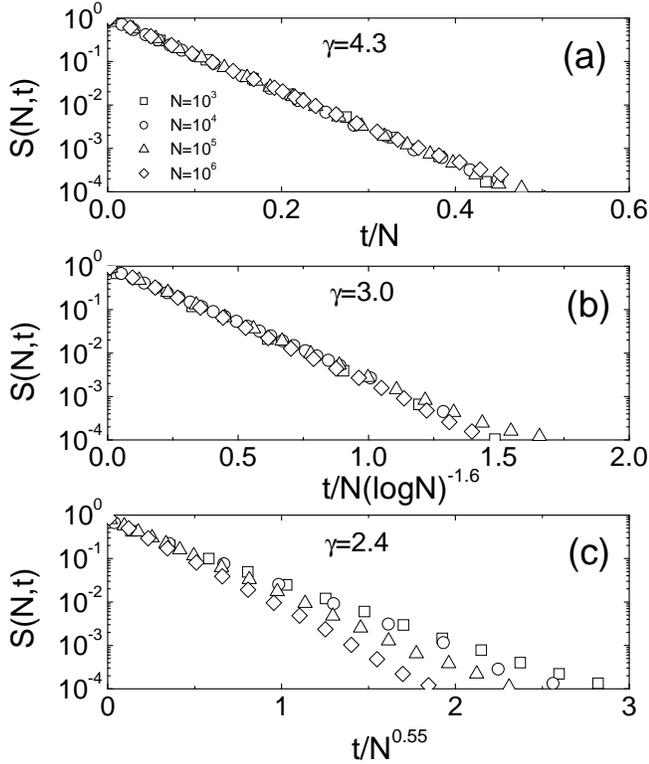}
\caption{The scaled plots of the survival probability $S(N,t)$ on
LSFNs with $\gamma=4.3$ (a), $3.0$ (b), and $2.4$ (c).
To scale the time $t$, $\alpha$'s are obtained from the fits of
$\left<T\right>\sim N^\alpha$ to the data (which are not shown
in this article).}
\label{SFN_ST}
\end{figure}

The dynamical behavior of one lamb and one lion model with the random
update rule on LSFNs has already been reported in Ref. \cite{cmaonsfn}.
We measure the $S(N,t)$ and $\left<T\right>$ for various $n$'s with
the random and the parallel update algorithms.
We check that $\left<T\right>$ and $S(N,t)$ scale 
in the same manner with those reported
in \REF{cmaonsfn}. Therefore, we
will report our results briefly in this section.
To generate the LSFNs of sizes $N=10^3 \sim 10^6$, we use the static
network model suggested by Ref. \cite{SNU_staticmodel}
with $\left<k\right>=4$.
In \FIG{SFN_ST} we display $S(N,t)$ for $n=10$ on LSFNs
with various $\gamma$(=4.3, 3.0, and 2.4).
$S(N,t)$ exponentially decays as
$S(N,t) \sim \exp(-t/\tau)$ for all $\gamma$'s.
As discussed in Sec. II, by assuming that $S(N,t)\sim \exp(-t/\tau)$,
we obtain the relation $\left<T\right>\sim \tau(N)$.
Using the value of $\alpha$'s obtained
from the best fit of $\left<T\right>\sim N^\alpha$ to the data (which
are not shown), we scale the time as $t/N$ for $\gamma>3$ and
as $t/N^{0.55}$ for $\gamma=2.4$.
As shown in \FIG{SFN_ST}(a), $S(N,t)$ for $\gamma> 3$ collapses
well into a single curve. This results implies that $\tau(N)\sim N$
for $\gamma>3$.
However, as we can see in \FIG{SFN_ST}(c), $S(N,t)$ for $\gamma=2.4$
does not collapse into a single curve by the simple scaling.
Note that $\gamma=3$ is marginal below which $\left<k^2\right>$
diverges. Thus, the logarithmic correction for the scaling
of $\tau$ can be expected for $\gamma=3$.
In \FIG{SFN_ST}(b), we scale the time as $t/N(\log N)^{-1.6}$
and find that $S(N,t)$ for $\gamma=3$ collapses into one curve.
This means that $\tau$ scales as $\tau\sim N(\log N)^{-1.6}$.
Therefore, we expect
\begin{equation}
\label{ST_SFN}
\lim_{t\rightarrow\infty}\lim_{N \rightarrow \infty} S(N,t)= \begin{cases}
S_o e^{-t/\tau_{\infty}} & (\gamma<3) \\
\mbox{finite} & (\gamma \ge 3)\;\;,
 \end{cases}
\end{equation}
where $\tau_{\infty}$ is the asymptotic value of $\tau$, and
depends on $\gamma$.
Eqs. (\ref{ST_TSFN}) and (\ref{ST_SFN}) show one of the
evidences that the dynamical behavior of the diffusive capture
process is related to $\left<k^2\right>$

\section{RELATION BETWEEN DEGREES AND CAPTURE EVENTS}

\begin{figure}
\includegraphics[width=8.5cm]{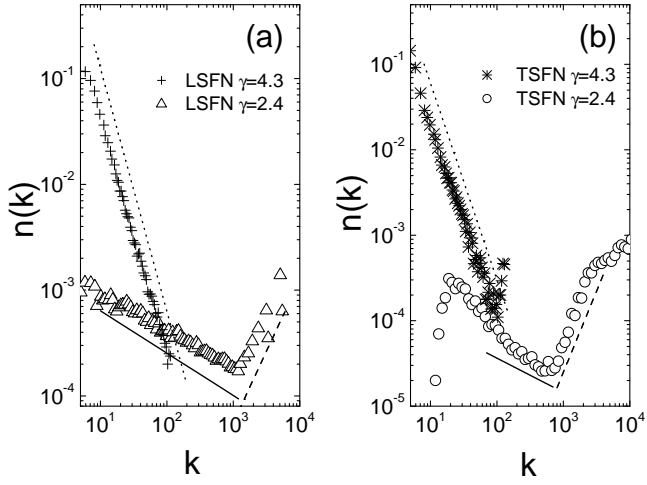}
\caption{
Plot of capture event distribution $n(k)$ on LSFN (a) and TSFN (b).
The lines stand for the relation $n(k)\sim k^{-\sigma}$.
The solid lines correspond to $\sigma=0.4$ and
the dotted lines represent $\sigma=2.3$.
The data for $k<10^3$ (a) and for $10^2<k<10^3$ (b) show good 
agreements with the relation $\sigma=\gamma-2$.
The dashed lines correspond to $\sigma=-1.35$ (a) and $\sigma=-1.9$ (b)
representing the anomalous increases of $n(k)$'s for large $k$.
}\label{nk}
\end{figure}

In the previous sections, we show that the underlying topology
significantly affects the dynamical behaviors of $S(N,t)$ and
$\left<T\right>$ of a lamb.
For the quantitative description of the
relationship between the structure of the networks and the dynamical
properties of lamb-lion model, we count the capture events
occurring at the nodes with degree $k$. \FIG{nk}(a) shows
the normalized number of capture events $n(k)$ on LSFNs with 1-lamb and
10-lions. As we can see, $n(k)$'s both on LSFNs and TSFNs with
$\gamma=4.3$ decay as $n(k)\sim k^{-\sigma}$ and the obtained $\sigma$
is $\sigma\simeq 2.3$.
However, $n(k)$ measured on LSFNs with
$\gamma=2.4$ decreases with $\sigma\simeq 0.4$ for $k<10^3$ and
increases with $\sigma\simeq -1.35$ for $k>10^3$.
On TSFNs with $\gamma=2.4$, we also find $\sigma\simeq 0.4$ for
$10^2<k<10^3$ and $\sigma\simeq -1.9$ for $k>10^3$ (see \FIG{nk}(b)).
The decreasing behavior of $n(k)$ for moderate $k$ can be explained by the
following analysis. In the steady-state, the probability that a
random walker visits a node $i$ with degree $k_i$ is given
by \cite{Noh92}
\bea
\label{P_i} P_i^\infty=A k_i,
\eea
where $A=1/\sum_i k_i$.
Therefore, the normalized capture events simply can be written as
\bea
\label{n_k}
n(k)= \sum_{i,j=1}^N P_i^\infty
\delta_{k_i,k} P_j^\infty \delta_{i,j} = A^2 k^2 NP(k).
\eea
If $P(k)\sim k^{-\gamma}$, then we obtain
\bea
\label{sigma1}
\sigma=\gamma-2.
\eea
Since we do not assume any topological property of underlying
networks, Eq. (\ref{sigma1}) is valid for any SFNs and 
the obtained values of $\sigma$ from the simulations for $k<10^3$
is consistent with \EQ{sigma1}.
Note that the $n(k)$'s anomalously increase for
$k>10^3$  both on the LSFNs and the TSFNs with $\gamma=2.4$.
This anomalous behavior of $n(k)$ can be explained by the
rarity of the dominant hubs in complex networks, i.e., 
the $P(k)$ for $k\approx k_{max}$ does not change much as $k$
varies.  Here $k_{max}$ is the degree of the biggest hub.
Thus, if we assume that $P(k\approx
k_{max})\sim const.$, then $n(k)$ increases as $n(k)\sim k^{2}$
or
\bea
\label{sigma2}
\sigma=-2,
\eea
for $k\approx k_{max}$. The obtained $\sigma(k\approx k_{max})$'s
for $\gamma=2.4$ from the numerical simulations are $-1.35$ on LSFN
and $-1.9$ on TSFN. The value of $\sigma(k\approx k_{max})$ obtained
from the simulation on LSFN with
$\gamma=2.4$ is considerably smaller than \EQ{sigma2}. This disagreement
might be originated from our assumption
that $P(k\approx k_{max})$ does not depend on $k$.
For example, if we assume that $P(k\approx k_{max}) \sim k^{-\epsilon}$
with $\epsilon<2$, then $\sigma$ becomes $\sigma=\epsilon-2$($<0$).
Eqs. (\ref{n_k}) and (\ref{sigma2}), thus, provide the
topological origin of the gathering behavior of random walkers at
the dominant hubs. It implies that the lamb and the lion have a
strong tendency to move into big hubs, i.e. the hubs collects the
lamb and the lion. Therefore, the hubs in the networks with strong
heterogeneity become the effective attractors of random walking lamb
and lions.

Another important consequence of \EQ{n_k} is that the dynamical
properties of lamb-lion model depend on $\left<k^2\right>$.
Note that from \EQ{n_k} the total number of capture event
$N_{tot}$ can be expressed as
\bea
\label{N_tot}
N_{tot}\propto\sum_{k} n(k)\sim\sum_{k} k^2 P(k)\sim
\left<k^2\right>.
\eea
If $\gamma<3$ then $N_{tot}$ also diverges.
This means that the lions on the networks can capture
the lamb no matter how the lamb is trying to escape
from the lions. Therefore, $S(N,t)\rightarrow 0$ in the limit $N\rightarrow
\infty$.
For $\gamma>3$, $N_{tot}$ converges to a finite value and
the lions can catch the lamb with a probability $p<1$,
thus $S(N,t)\ne 0$ for large $N$.
The expectations agree well with our simulation results
shown in Sec. II and III.
Based on \EQ{N_tot}, we also expect that the long-living
tail of $S(N,t)$ on TSFN with $\gamma<3$ eventually goes to 0.

\section{SUMMARY and discussion}

We investigate the diffusive capture process on two different
kinds of SFNs. The diffusive capture process can be mapped into
the diffusing lamb and lion model. We measure $S(N,t)$ and
$\left<T\right>$ of a lamb.
We find that on both TSFNs and LSFNs $S(N,t)$ of a lamb
decays exponentially for $\gamma> 3$. Simple scaling
arguments predict that the exponentially decaying
$S(N,t)$ for $\gamma> 3$ is originated from the finite size effect.
This result recovers the mean-field expectation
$S(N,t)\rightarrow finite$ in the limit $N\rightarrow \infty$.
However, we find that $S(N,t)$ on TSFNs with $\gamma<3$ has
the long-living tail. By measuring
the number of capture events with life time $T$ we find that
the long-living tail is caused by enormously many short branches in TSFNs.
%Note that $\gamma=3$ is marginal, and we find a logarithmic
%correction in scaling of $\tau(N)$ of $S(N,t)$.
On LSFN with $\gamma<3$, we find that $S(N,t)$ decays
exponentially and no lamb can survive in thermodynamic limit.

In order to understand the non-mean-field behavior of $S(N,t)$
for $\gamma<3$, we study the normalized number of capture events
$n(k)$ both numerically and analytically.
From the numerical simulations, we find that $n(k)$ decreases as
a power-law, $n(k)\sim k^{-\sigma}$.
The simple analytic derivation of $n(k)$ provides the relation
$\sigma=\gamma-2$, which shows a good agreement with numerical
results for moderate $k$.
However, we find that $n(k)$ anomalously increases for large $k$
in the simulations. This means that the most of the lambs are captured
by lions at the hubs. Hence, those hubs become the effective attractors.
By assuming the rarity of dominant hubs and $P(k)\approx const.$
for large $k$($\approx k_{max}$), we show that the $n(k)$
can increase for large $k$.
From the analytic expectation of $n(k)$, we also show that $N_{tot}\sim
\left<k^2\right>$.
This implies that if the fluctuation in degree is large enough
($\gamma<3$) then the dynamical behaviors of the diffusive capture
process deviate from the mean-field expectations by emerging effective
attractors of random walkers.
From the analytic expectation of $n(k)$ and $N_{tot}$,
we also expect that the long-living tail of $S(N,t)$ observed on
TSFN with $\gamma<3$ (see \EQ{ST_TSFN}) will eventually go to 0
due to the diverging $\left<k^2\right>$.

This work is supported by Korea Research Foundation
Grant No. KRF-2004-015-C00185.

\newpage

\end{document}